\documentclass{article}
\usepackage[T1]{fontenc}
\usepackage{wrapfig}
\usepackage[portuges,english]{babel}
\usepackage{amssymb,latexsym,amsmath,color,mathrsfs,ifsym,graphics,stmaryrd} 
\usepackage[colorlinks,linkcolor=blue,urlcolor=blue,citecolor=black,
plainpages=false,pdfpagelabels,breaklinks]{hyperref}

\title{Unscrambling the  Omelette of Quantum Contextuality (Part I):\\ Preexistent Properties or Measurement Outcomes?}

\author{{\sc Christian de Ronde}\thanks{Fellow Researcher of the Consejo
Nacional de Investigaciones Cient\'{\i}ficas y T\'ecnicas. Philosophy Institute ``Dr. A. Korn'', Buenos Aires University. Adjoint Profesor of the National University Arturo Jauretche. E-mail: cderonde@vub.ac.be}}
\date{}

\usepackage[margin=2.5cm]{geometry}

\begin{document}
\maketitle

\begin{center}
\begin{small}
CONICET, Buenos Aires University - Argentina \\
Center Leo Apostel and Foundations of  the Exact Sciences\\
Brussels Free University - Belgium \\
\end{small}
\end{center}

\begin{abstract}
\noindent In this paper we attempt to analyze the physical and philosophical meaning of quantum contextuality. We will argue that there exists a general confusion within the foundational literature arising from the improper ``scrambling'' of two different meanings of {\it quantum contextuality}. While the first one, introduced by Bohr, is related to an {\it epistemic} interpretation of contextuality which stresses the incompatibility (or complementarity) of certain measurement situations described in classical terms; the second meaning of contextuality is related to a purely {\it formal} understanding of contextuality as exposed by the Kochen-Specker (KS) theorem which focuses instead on the constraints of the orthodox quantum formalism in order to interpret projection operators as preexistent or actual (definite valued) properties. We will show how these two notions have been scrambled together creating an ``omelette of contextuality'' which has been fully widespread through a popularized ``epistemic explanation'' of the KS theorem according to which: {\it The measurement outcome of the observable $A$ when measured together with $B$ or together with $C$ will necessarily differ in case $[A, B] = [A, C] = 0$, and $[B, C] \neq 0$.}  We will show why this statement is not only improperly scrambling epistemic and formal perspectives, but is also physically and philosophically meaningless. Finally, we analyze the consequences of such widespread epistemic reading of KS theorem as related to statistical statements of measurement outcomes.
\end{abstract}

\newpage
\renewenvironment{enumerate}{\begin{list}{}{\rm \labelwidth 0mm
\leftmargin 0mm}} {\end{list}}

\newcommand{\ita}{\textit}
\newcommand{\mcal}{\mathcal}
\newcommand{\mfrak}{\mathfrak}
\newcommand{\mbb}{\mathbb}
\newcommand{\mrm}{\mathrm}
\newcommand{\msf}{\mathsf}
\newcommand{\mscr}{\mathscr}
\newcommand{\lra}{\leftrightarrow}
\renewenvironment{enumerate}{\begin{list}{}{\rm \labelwidth 0mm
\leftmargin 5mm}} {\end{list}}

\newtheorem{dfn}{\sc{Definition}}[section]
\newtheorem{thm}{\sc{Theorem}}[section]
\newtheorem{lem}{\sc{Lemma}}[section]
\newtheorem{cor}[thm]{\sc{Corollary}}
\newcommand{\Proof}{\textit{Proof:} \,}
\newcommand{\cqd}{{\rule{.70ex}{2ex}} \medskip}

\section*{Introduction}

Quantum contextuality is not only important for the philosophical understanding of Quantum Mechanics (QM), but is also key to the present technological developments of quantum information processing. This paper attempts to shed some light on the physical and philosophical meaning of quantum contextuality by discussing the following statements:

\begin{enumerate}
\item[1.] There are two different notions of contextuality in QM confused within the foundational literature. On the one hand, an epistemic notion which discusses about (classical) measurement processes and outcomes, and on the other, an ontic notion of contextuality which discusses in purely formal-conceptual terms about the interpretation of projection operators as preexistent ---meaning, independent of measurement observations--- properties.
\item[2.] KS theorem provides formal constraints to interpret projection operators as preexistent properties making no reference whatsoever to measurement outcomes.\footnote{As we shall see, contrary to a widespread presupposition, the theorem does not relate to ``classical experimental situations'', such as those discussed repeatedly by Bohr (e.g., \cite{Bohr35}).} 
\item[3.] Peres' work has created the illusion that it is possible to interpret the KS theorem ---following the epistemic viewpoint--- in terms of measurement outcomes (see e.g., \cite{Peres91, Peres02}). 
\item[4.] Peres interpretation might be also responsible for the illusion that KS theorem can be easily related to measurable statistical inequalities \cite{Cabello98, Cabello08, KunjwalSpekkens15, Spekkens05}. We will discuss why such inequalities are not directly related to the KS theorem.
\end{enumerate}
 
Our analysis attempts to highlight the deep importance of philosophical considerations for the proper elucidation of foundational debates about QM. It also goes in line with David Mermin's intuition according to which ``the whole notion of an experimental test of KS [theorem] misses the point''.\footnote{Mermin as cited by A. Cabello in \cite{Cabello98}.} The paper is organized as follows. In the first section we discuss the main differences between epistemic and ontic views within the debate about the meaning of quantum theory. In section 2 we address the omelette that has been created within the foundational literature regarding objective and subjective perspectives, ontological and epistemlogical, etc. Section 3 presents the epistemic understanding of contextuality grounded on Bohr's analysis of complementary (or incompatible) measurement situations. In section 4 we provide an ontic definition of contextuality grounded on KS theorem which makes reference to preexistent properties ---and no reference whatsoever to measurement outcomes. In section 5 we criticize Peres' epistemic interpretation of KS theorem in terms of measurement outcomes. Finally, in section 6, we analyze the consequences of such widespread epistemic reading of KS theorem as related to statistical statements of measurement outcomes. We end our paper with a short conclusion.

\section{Epistemic vs Ontic Views in QM}

The epistemic and ontic views have played a significant role in the debate about the meaning and interpretation of QM. As we shall argue, the improper scrambling of these two distinct philosophical perspectives have also played a fundamental role in the misunderstanding of one of the main characteristics of the theory of quanta, known in the literature as quantum contextuality. In order to address this notion we shall begin our paper by a short review of the role played by both epistemic and ontic viewpoints within the theory of quanta. Our intention is not to provide an historic account of contextuality nor to discuss the ``true interpretation'' of Bhorian philosophy of physics, but rather to situate quantum contextuality within the more general debate about the philosophical meaning of QM. We should also stress that even though the definitions of what is to be considered an `epistemic view' and what is `ontic view' are obviously debatable, there are some general characteristics which we will attempt to differentiate. Our main interest is to distinguish between two very different approaches regarding the understanding of QM.   

According to Marchildon \cite{Marchildon04} ``In the epistemic view, the state vector (or wave function, or density matrix) does not represent the objective state of a microscopic system (like an atom, an electron, a photon), but rather our knowledge of the probabilities of outcomes of macroscopic measurements.'' This perspective is not new in QM. It might be traced back to Niels Bohr's interpretation of the theory as a rational generalization of classical mechanics \cite{BokulichBokulich}. Indeed, the approach of the Danish physicist might be regarded as one of the first attempts to understand QM in epistemological ---rather than ontological--- terms. Bohr stressed repeatedly that the most important lesson we should learn from QM is an epistemological one; namely, that we are not only spectators, but also actors in the great drama of existence. This idea goes in line with his pragmatic understanding of physics itself according to which \cite{Bohr60}: ``Physics is to be regarded not so much as the study of something a priori given, but rather as the development of methods of ordering and surveying human experience. In this respect our task must be to account for such experience in a manner independent of individual subjective judgement and therefor objective in the sense that it can be unambiguously communicated in ordinary human language.'' The epistemic interpretation of QM was developed in many different directions and has become a standard way of understanding the theory, one which attempts to evade the metaphysical problems and questions discussed by realists. Indeed, as Marchildon remarks [{\it Op. cit.}], it is argued by its followers that the epistemic viewpoint ``considerably clarifies, or even completely dissolves, the EPR paradox and the measurement problem.'' 

In the present, the epistemic viewpoint has been taken to its most radical limit by an approach put forward by Christopher Fuchs, Asher Peres, R\"udiger Schack and David Mermin, known as ``Quantum Bayesianism'', or in short: QBism. This might be regarded as one of the most radical epistemic approaches and becomes a good example of what this perspective amounts to. Let us discuss it in some detail. QBism argues explicitly against an ontological reading of QM and assumes a subjectivist perspective taking as a standpoint the epistemic Bayesian understanding of probability \cite{QBism13}.\footnote{To be more precise, even though the epistemic view has many different proponents, we believed that QBism was the most consistent, clear and honest epistemic account of QM (see for a detailed analysis \cite{deRonde16e}). This was until Fuchs and Mermin begun to argue that QBims is also consistent with some weird kind of ``subjective realism'', abandoning in this way Peres' original anti-realist viewpoint. Either `QM talks about physical reality' or `QM does not talk about physical reality'. Both cannot be true simultaneously. These two contradictory statements are scrambled together in \cite{Fuchs16, Mermin14b}.} The origin of QBism might be traced back to a paper published in the year 2000 entitled {\it Quantum Theory Needs no `Interpretation'}. In that paper, Fuchs and Peres \cite[p. 70]{FuchsPeres00} argued that: ``[...] quantum theory does not describe physical reality. What it does is provide an algorithm for computing probabilities for the macroscopic events (`detector clicks') that are the consequences of experimental interventions. This strict definition of the scope of quantum theory is the only interpretation ever needed, whether by experimenters or theorists.'' As remarked by Fuchs, Mermin and Shack \cite[p. 2]{QBism13}: ``QBism explicitly takes the `subjective' or `judgmental' or `personalist' view of probability, which, though common among contemporary statisticians and economists, is still rare among physicists: probabilities are assigned to an event by an agent and are particular to that agent. The agent's probability assignments express her own personal degrees of belief about the event.'' According to them: ``A measurement in QBism is more than a procedure in a laboratory. It is any action an agent takes to elicit a set of possible experiences. The measurement outcome is the particular experience of that agent elicited in this way. Given a measurement outcome, the quantum formalism guides the agent in updating her probabilities for subsequent measurements.'' QBists are crystal clear: ``A measurement does not, as the term unfortunately suggests, reveal a pre-existing state of affairs.'' Measurements are ``personal'' and QM is a ``tool'' for the ``user''  ---as Mermin prefers to call the ``agent'' \cite{Mermin14}. 

The focus of the epistemic viewpoint in the measurement processes and the observation of outcomes is clearly confronted by the ontic viewpoint which insists on interpreting QM beyond observable measurement outcomes. The ontic perspective, in the context of QM, can be obviously related to Albert Einstein's philosophical position which confronted Bohr's epistemic understanding of physics. According to this view, it is the conceptual representation provided by a theory that which expresses ---in some way--- what reality is about ---completely independently of human choices and conscious beings. Let us remark that the specific way in which theoretical representation relates to reality is not something obvious nor ``self evident''. Realism is not the na\"ive belief ---supported by scientific realism--- that theories describe in a one to one manner {\it reality as it is} (see \cite{deRonde17}).\footnote{That would imply a Platonist heaven of ``true concepts'' waiting still to be discovered.} It is neither the belief of a human subject in the existence of an external reality. The belief that `the world is out there' is simply not enough to characterize what realism is about. Realism is a program which understands ---since the Greeks--- that {\it theories} can capture in some way {\it physis} ---nature or reality. The way in which this is done varies from one particular metaphysical perspective to the other. However, without a link between reality and theoretical representation it is not possible to articulate realism. Realism attempts to provide both a theoretical representation and the link to {\it physis}. Of course, there are many possibilities to address the relation between reality and theoretical representation, one might investigate different types of transcendent or immanent schemes.\footnote{In \cite{deRonde14} we have proposed a neo-Spinozist pluralist scheme based on Heisenberg's closed theories approach which understands the representational character of physical theories as immanent expressions ---rather than descriptions--- of {\it physis}. A different transcendent non-reductionistic proposal which might be considered is the {\it Multiplex Realism} proposed by Aerts and Sassoli de Bianchi in \cite{AertsSassoli15}.} However, regardless of the choice, the realist is always stuck with these problems; i.e. trying to find adequate physical representations and trying to explain their specific relation to {\it physis}. 

A deeply important point in this respect is that one cannot be a realist about observations. Observability is considered by the realist as a necessarily derived theoretical notion. As remarked by Einstein \cite[p. 175]{Dieks88a}: ``[...] it is the purpose of theoretical physics to achieve understanding of physical reality which exists independently of the observer, and for which the distinction between `direct observable' and `not directly observable' has no ontological significance.'' Unlike positivism, Einstein did not accept na\"ive empiricism; i.e. the consideration of observables as ``self evident'' {\it givens}.\footnote{This can be seen from the very interesting discussion between Heisenberg and Einstein \cite[p. 66]{Heis71} were the latter explains: ``I have no wish to appear as an advocate of a naive form of realism; I know that these are very difficult questions, but then I consider Mach's concept of observation also much too naive. He pretends that we know perfectly well what the word `observe' means, and thinks this exempts him from having to discriminate between `objective' and `subjective' phenomena.''} For Einstein, the interrelation between physics and metaphysics was a central element of physics itself (see \cite{Howard94}). He also stressed, like Heisenberg, the importance of developing new physical concepts in order to adequately address new fields of phenomena, new experiences. It is only the formalism of a theory adequately related to physical concepts which is capable of describing phenomena. As Einstein remarked to a young Heisenberg \cite[p. 63]{Heis71}: ``It is only the theory which decides what we can observe.'' Heisenberg \cite[p. 264]{Heis73} himself would later emphasize this same point:  ``The history of physics is not only a sequence of experimental discoveries and observations, followed by their mathematical description; it is also a history of concepts.  For an understanding of the phenomena the first condition is the introduction of adequate concepts. Only with the help of correct concepts can we really know what has been observed.''

To summarize, the epistemic and the ontic viewpoints face right from the start very different problems and concerns. While the epistemic view concentrates in the way subjects (also called `agents', `users', etc.) are capable of relating to observable measurement outcomes, the ontic perspective focuses on the physical meaning and interpretation of the formalism of a theory ---independently of individual subjects. In this case, observability ---as Einstein himself stressed repeatedly---  has no ontological significance; it is only regarded as part of a verification procedure about the empirical adequacy (or not) of the theory in question. While the epistemic proponent might understand the theory as a mere ``algorithm'' or ``economy'' of measurement outcomes, for the ontic viewer a theory and its formalism are telling us something about physical reality ---the problem is how to represent what the theory is telling us in conceptual terms. While the former takes the observability of subjects to be the starting point of science and also the end, the latter fundaments the theory in relation to the objective ---i.e., subject independent--- representation of physical reality. Unfortunately, within QM, these two radically opposite viewpoints ---in what respects the presuppositions involved in the meaning of a theory--- have been mixed in a ``omelette'' that is still today being cooked.

\section{The Quantum Omelette}

The philosophical stance that we assume defines the specific problems, the possible questions and (even) answers that fall within our system of thought. But a limit is also a possibility, an horizon. A particular perspective limits the possibility of questioning but it also constitutes it. Problems are not ``out there'' independent of philosophical perspectives, they are part of a definite viewpoints with particular metaphysical assumptions, presuppositions, without which problems could not be even stated. For example, a reductionistic perspective compared to a pluralist one will radically differ when considering the quantum to classical limit \cite{deRonde16b}. And while a realist considers as necessary the existence of an objective representation which explains what the formalism is talking about, the instrumentalist might remain content with an intersubjective epistemic account of measurement outcomes operationally provided {\it via} a mathematical formalism understood merely as an algorithm. While the problem for the realist will be to provide a conceptual representation that matches the mathematical formalism, the problem for the instrumentalist will be to explain how is it possible to bridge the gap between a purely mathematical formal scheme and actual observations.\footnote{Regardless of the fact it s claimed today that this can be done very easily by providing a set of {\it interpretational rules}, this problem has not been yet properly addressed nor solved. In fact, this question has hunted philosophy of physics since it positivist origin. Today, instead of trying to address this still unsolved problem, in the present literature a form of naive empiricism seems to be the uncritical basis of many foundational debates.} In this respect, the situation within the philosophy of QM, has become completely unacceptable. Today, we are at a stage where ontology and epistemology, objectivity and subjectivity have been mixed up in an ``omelette'' that we urgently need to unscramble. As Jaynes made the point:

\begin{quotation}
\noindent {\small``[O]ur present [quantum mechanical] formalism is not purely epistemological; it is a peculiar mixture describing in part realities of Nature, in part incomplete human information about Nature ---all scrambled up by Heisenberg and Bohr into an omelette that nobody has seen how to unscramble. Yet we think that the unscrambling is a prerequisite for any further advance in basic physical theory. For, if we cannot separate the subjective and objective aspects of the formalism, we cannot know what we are talking about; it is just that simple.''  \cite[p. 381]{Jaynes}}
\end{quotation}

It is interesting to notice, however, that one of the first scramblings of objective and subjective was produced, neither by Bohr nor by Heisenberg, but from one of the strongest attacks to the theory of quanta. In the famous EPR paper \cite{EPR}, the definition of what had to be considered an {\it element of physical reality} ---in the context of QM--- begun the explicit scrambling between an objective {\it ontic} definition of reality and  the subjective, {\it epistemic} reference, to the quantum measurement process. According to the famous definition: {\it  If, without in any way disturbing a system, we can predict with certainty (i.e., with probability equal to unity) the value of a physical quantity, then there exists an element of reality corresponding to that quantity.} As it is known in the foundational literature, the problem comes with the phrase: ``without in any way disturbing the system''.\footnote{See in this respect the famous paper by Bell: {\it Against `Measurement'} \cite{Bell}.} Bohr himself stressed this fact, in his also famous reply \cite[p. 697]{Bohr35}, arguing that the ``criterion of physical reality [...] contains ---however cautions its formulation may appear--- an essential ambiguity when it is applied to the actual problems with which we are here concerned.'' Bohr used this ``ambiguity'' to reintroduce the epistemic analysis of measurements considering  ``in some detail a few simple examples of measuring arrangements'' [{\it Op. cit}, p. 697]. In this way he was able to shift the debate from ontology to epistemology, leaving the question posed by Einstein completely unanswered. 

Einstein knew very well that a realist should not define physical reality in epistemic terms for that would involve the improper intromission of a subject, breaking down the very possibility to consider an `objective reality' detached from subjective choices. It was his student Podolsky who had come up with this ambiguous definition that Einstein disliked and even criticized in conversations with Schr\"odinger. Bohr, as a Kung Fu master, immediately detected this weakness and used the attack of the opponent to turn it against himself. In his brilliant reply, he deconstructed EPR's whole argumentation by concentrating on the measurement process alone. Bohr was not willing to enter the truly ontological debate ---regarding physical representation--- proposed by Einstein (see for a detailed discussion \cite{Bacc14}). As remarked by his assistant Aage Petersen:  

\begin{quotation}
\noindent {\small ``Traditional philosophy has accustomed us to regard
language as something secondary and reality as something primary.
Bohr considered this attitude toward the relation between language
and reality inappropriate. When one said to him that it cannot be
language which is fundamental, but that it must be reality which, so
to speak, lies beneath language, and of which language is a picture,
he would reply, ``We are suspended in language in such a way that we
cannot say what is up and what is down. The word `reality' is also a
word, a word which we must learn to use correctly'' Bohr was not
puzzled by ontological problems or by questions as to how concepts
are related to reality. Such questions seemed sterile to him. He saw
the problem of knowledge in a different light.'' \cite[p. 11]{Petersen63}}
\end{quotation}

Bohr was very careful not to enter ontic debates and always remained within the limits of his epistemic analysis. But Heisenberg was not so careful. He wanted, on the one hand, to support Bohr's epistemological approach, and on the other, his own Platonist ontological scheme. In his book, {\it Physics and Philosophy} \cite{Heis58}, Heisenberg begun to create what is known today as ``The Copenhagen Interpretation of QM'' (see for discussion \cite{Howard04}). This interpretation attempted to bring together not only Bohr's epistemological approach and the necessity of classical language but also Heisenberg's own Platonist realism about mathematical equations in physical theories \cite[p. 91]{Heis71}. While Bohr's anti-metaphysical commitment considered the language of Newton and Maxwell as the fundament of all possible physical phenomena, Heisenberg's closed theory approach insisted in the radical incommensurability of the physical concepts used by different theories \cite{Bokulich04}. This book shows not only Heisenberg's fantastic historical and philosophical knowledge about physics, but also the omelette he created scrambling improperly objective and subjective elements, ontology and epistemology. A good example of the quantum omelette cooked in Heisenberg's book is the following passage:

\begin{quotation}
\noindent {\small``With regard to this situation Bohr has emphasized that it is more realistic to state that the division into the object and the rest of the world is not arbitrary. Our actual situation in research work in atomic physics is usually this: we wish to understand a certain phenomenon, we wish to recognize how this phenomenon follows from the general laws of nature. Therefore, that part of matter or radiation which takes part in the phenomenon is the natural `object' in the theoretical treatment and should be separated in this respect from the tools used to study the phenomenon. {\it This again emphasizes a subjective element in the description of atomic events, since the measuring device has been constructed by the observer, and we have to remember that what we observe is not nature in itself but nature exposed to our method of questioning.} Our scientific work in physics consists in asking questions about nature in the language that we possess and trying to get an answer from experiment by the means that are at our disposal. In this way quantum theory reminds us, as Bohr has put it, of the old wisdom that when searching for harmony in life one must never forget that in the drama of existence we are ourselves both players and spectators. It is understandable that in our scientific relation to nature our own activity becomes very important when we have to deal with parts of nature into which we can penetrate only by using the most elaborate tools.'' \cite[p. 9]{Heis58} (emphasis added)}
\end{quotation}

\noindent The confusion comes from the fact that acknowledging physical representation and experience have been created through concepts and tools specifically designed by us, humans, is completely different from claiming that the choice of a specific experimental arrangement determines physical reality itself. In Kantian terms, this is the confusion between the subjective transcendental scheme (i.e., the categorical conditions of objectivity) which allows for the possibility of objective representation and particular empirical subjects (i.e., individual subjects, agents or users) which are in fact part of the categorical representation itself.\footnote{For a detailed analysis of this important distinction we refer to \cite{deRonde17}.} This situation in QM, regarding the subjective definition of physical reality, was clearly recognized by Einstein who remained always uncomfortable with the epistemological reasoning of Bohr. As recalled by Pauli:

\begin{quotation}
\noindent {\small ``{\it Einstein}'s opposition to [quantum mechanics] is again reflected in his papers which he published, at first in collaboration with {\small \emph{Rosen}} and {\small \emph{Podolsky}}, and later alone, as a critique of the concept of reality in quantum mechanics. We often discussed these questions together, and I invariably profited very greatly even when I could not agree with {\small \emph{Einstein}}'s view. `Physics is after all the description of reality' he said to me, continuing, with a sarcastic glance in my direction `or should I perhaps say physics is the description of what one merely imagines?' This question clearly shows {\small \emph{Einstein}}'s concern that the objective character of physics might be lost through a theory of the type of quantum mechanics, in that as a consequence of a wider conception of the objectivity of an explanation of nature the difference between physical reality and dream or hallucination might become blurred.'' \cite[p. 122]{Pauli94}}
\end{quotation}

Independently of Einstein's efforts to discuss the possible physical representation of quantum reality, the Bohrian approach has become a silent orthodoxy that limits the analysis of QM down to  the almost exclusive set of problems which ---following Bohr's interpretation--- presuppose the representation of reality that results from classical physics. In the following section we attempt to discuss the specific notion of contextuality that Bohr himself derived following his own philosophical understanding of physics.

\section{Bohrian (Epistemic) Contextuality}

Bohr's notion of contextuality is directly related to his solution of the wave-particle duality in the double-slit experiment and the necessary requirement to provide an account of each experimental set up in terms of classical concepts.\footnote{See for a detailed analysis: \cite{BokulichBokulich}.} The epistemic solution presented by Bohr was provided through his notion of complementarity which allowed him to consider mutually incompatible classical representations such as those of `wave' and `particle'. 

According to Bohr, the problem arises when attempting to distinguish what is the observed system and what the apparatus, what plays the role of the subject and what of the object.  

\begin{quotation}
\noindent {\small``This necessity of discriminating in each experimental arrangement between those parts of the physical system considered which are to be treated as measuring instruments and those which constitute the objects under investigation may indeed be said to form a {\it principal distinction between classical and quantum-mechanical description of physical phenomena.}'' [{\it Op. Cit.}, p. 701] (emphasis in the original)}
\end{quotation}

\noindent As stressed by Bohr [{\it Op. Cit.}], this fundamental distinction ``has its root in the indispensable use of classical concepts in the interpretation of all proper measurements, even though the classical theories do not suffice in accounting for the new types of regularities with which we are concerned in atomic physics.'' Bohr's contextuality implied the need to discuss in terms of `classical contexts'; i.e. contexts as related to classical experimental apparatuses described in terms of classical physics. This (metaphysical) requirement was not justified by the quantum formalism, but rather by his insistence in the necessity of committing to the use of classical physical language. Bohr \cite[p. 7]{WZ} stated that: ``[...] the unambiguous interpretation  of any measurement must be essentially framed in terms of classical physical theories, and we may say that in this sense the language of Newton and Maxwell will remain the language of physicists for all time.'' In this respect, he argued ``it would be a misconception to believe that the difficulties of the atomic theory may be evaded by eventually replacing the concepts of classical physics by new conceptual forms.''\footnote{This is in fact what David Deutsch \cite{Deutsch11} has characterized as ``bad philosophy'', namely,  ``philosophy that is not merely false, but actively prevents the growth of other knowledge.'' He then continues his explanation with a direct attack to the Bohrian philosophy: ``The physicist Niels Bohr (another of the pioneers of quantum theory) then developed an `interpretation' of the theory which later became known as the `Copenhagen interpretation'. It said that quantum theory, including the rule of thumb, was a complete description of reality. Bohr excused the various contradictions and gaps by using a combination of instrumentalism and studied ambiguity. He denied the `possibility of speaking of phenomena as existing objectively' ---but said that only the outcomes of observations should count as phenomena. He also said that, although observation has no access to `the real essence of phenomena', it does reveal relationships between them, and that, in addition, quantum theory blurs the distinction between observer and observed. As for what would happen if one observer performed a quantum-level observation on another, he avoided the issue ---which became known as the `paradox of Wigner's friend', after the physicist Eugene Wigner.''} This metaphysical precondition was in line with his neo-Kantian philosophical perspective according to which phenomena must be necessarily considered as classical space-time phenomena. In this respect, it is important to stress that Bohr never discussed in terms of the mathematical formalism of the theory, which he considered to be a ``purely symbolic scheme''. Instead, he took as a standpoint the representation of experimental set ups in terms of classical physics and its language. Bohr assumed the epistemic perspective according to which a subject must always describe experimental situations in terms of classical physics. In this sense, he was focused ---as it can be clearly seen in his reply to EPR--- in the process of measurement (see for discussion \cite{Bacc14}). We might comprise his definition of contextuality in the following terms:

\begin{dfn} 
{\bf Bohr's (Epistemic) Contextuality:} Contexts, which are considered in terms of experimental situations described in terms of classical physical concepts, are incompatible in QM. The most paradigmatic example of this notion of contextuality is the wave-particle duality present in the double-slit experiment.
\end{dfn}

\noindent According to this understanding the effect of the measurement apparatus is to create a value (of an observable), which did not exist before. This is perfectly consistent with Bohr's remark that the most important lesson of QM was and epistemic one, namely, that we are not only spectators but also actors in the great drama of (quantum) existence. As John Bell himself puts it when analyzing Bohr's reply to EPR:

\begin{quotation}
\noindent {\small ``The result of a `spin measurement', for example, depends in a very complicated way on the initial position {\it x} of the particle and on the strength and geometry of the magnetic field. Thus the result of the measurement does not actually tell us about some property previously possessed by the system, but about something which has come into being in the combination of system and apparatus.''
\cite[p. 35]{Bell87}}
\end{quotation}

\noindent ``Very complicated'' means here that there is no physical explanation nor representation of the process, and thus one must content oneself with the belief that ``quantum particles'' ---whatever they are--- are able in some strange ``magical'' way ---which has not been explained up to the present--- to transform themselves into classical systems. 

Regardless of the acceptance or not of such perspective ---widespread in the orthodox literature--- we would like to stress that this epistemic definition of contextuality implies metaphysical presuppositions which have no direct relation to the orthodox formalism of QM. The idea that ``all experience must be described in classical terms'' is not an ``obvious'' nor ``self evident'' statement. In fact, there is no constraint of the formalism regarding the possibility to understand and discuss quantum contextuality beyond classical concepts ---a possibility which Bohr considered inadmisible. This ideas have also sedimented the widespread claim according to which contextuality must be understood as the fact that in QM, ``the properties of a system are different whether you look at them or not'' \cite{Butterfield17}. 

Contrary to the epistemic viewpoint, we believe that the development of a non-classical representation of QM remains a logical possibility that should not be erased from the field of research without good reasons. As stressed by Einstein: 

\begin{quotation}
\noindent {\small ``Concepts that have proven useful in ordering things easily achieve such an authority over us that we forget their earthly origins and accept them as unalterable givens. Thus they come to be stamped as `necessities of thought,' `a priori givens,' etc. The path of scientific advance is often made impossible for a long time  through such errors. For that reason, it is by no means an idle game if we become practiced in analyzing the long common place concepts and exhibiting those circumstances upon which their justification and usefulness depend, how they have grown up, individually, out of the givens of experience. By this means, their all-too-great authority will be broken. They will be removed if they cannot be properly legitimated, corrected if their correlation with given things be far too superfluous, replaced by others if a new system can be established that we prefer for whatever reason.'' \cite{Howard10}}
\end{quotation}

\section{Kochen-Specker (Formal-Ontic) Contextuality}

Contrary to Bohr, Simon Kochen and Ernst Specker begun their analysis, taking as a standpoint the orthodox formalism of QM, asking ---implicitly--- a question related to the possible ontology of the theory of quanta which has no epistemic reference whatsoever. {\it Would it be possible to consider projection operators as actual (definite valued) preexistent properties within the orthodox formalism of QM?}  This question, implicit in their analysis, led them to a very interesting result which we now shortly recall. 

In QM the frames under which a vector is represented mathematically are considered in terms of orthonormal bases. We say that a set $\{x_1,\ldots,x_n\}\subseteq {\cal H}$ an $n$-dimensional Hilbert space is an \emph{orthonormal basis} if $\langle x_{i} | x_{j} \rangle = 0$ for all $1 \leq i , j \leq n$ and $\langle x_i|x_i\rangle=1$ for all $i=1,\ldots,n$. A physical quantity is represented by a self-adjoint operator on the Hilbert space ${\cal H}$. We say that $\mathcal{A}$ is a $\emph{context}$ if $\mathcal{A}$ is a commutative subalgebra generated by a set of self-adjoint bounded operators $\{A_1,\ldots,A_s\}$ of ${\cal H}$. Quantum contextuality, which was most explicitly recognized through the Kochen-Specker (KS) theorem \cite{KS}, asserts that a value ascribed to a physical quantity $A$ cannot be part of a global assignment of values but must, instead, depend on some specific context from which $A$ is to be considered. Let us see this with some more detail.

Physically, a global valuation allows us to define the preexistence of definite properties. Mathematically, a  \emph{valuation} over an algebra $\mathcal{A}$ of self-adjoint operators on a Hilbert space, is a real function satisfying,

\begin{enumerate}
\item[1.] \emph{Value-Rule (VR)}: For any $A\in\mathcal{A}$, the value $v(A)$ belongs to the spectrum of $A$, $v(A)\in\sigma(A)$.
\item[2.] \emph{Functional Composition Principle (FUNC)}: For any $A\in\mathcal{A}$ and any real-valued function $f$, i.e. $v(f(A))=f(v(A))$.
\end{enumerate}

\noindent We say that the valuation is a \emph{Global Valuation (GV)} if $\mathcal{A}$ is the set of all bounded, self-adjoint operators. In case $\mathcal{A}$ is a context, we say that the valuation is a \emph{Local Valuation (LV)}. We call the mathematical property which allows us to paste consistently together multiple contexts of {\it LVs} into a single {\it GV}, {\it Value Invariance (VI)}. First assume that a {\it GV} $v$ exists and consider a family of contexts $\{ A_i \}_I$. Define the {\it LV} $v_i:=v|_{A_i}$ over each $A_i$. Then it is easy to verify that the set $\{v_i\}_I$ satisfies the \emph{Compatibility Condition (CC)}, 

$$v_i|_{ A_{i} \cap A_j} =v_j|_{A_i\cap A_j},\quad \forall i,j\in I.$$

\noindent The {\it CC} is a necessary condition that must satisfy a family of {\it LVs} in order to determine a {\it GV}. We say that the algebra of self-adjoint operators is \emph{VI} if for every family of contexts $\{ A_i\}_I$ and {\it LVs} $v_i: A_i \rightarrow \mathbb{R}$ satisfying the \emph{CC}, there exists a {\it GV} $v$ such that $v|_{A_i}=v_i$.

If we have {\it VI}, and hence, a {\it GV} exists, this would allow us to give values to all magnitudes at the same time maintaining a {\it CC} in the sense that whenever two magnitudes share one or more projectors, the values assigned to those projectors are the same in every context. The KS theorem, in algebraic terms, rules out the existence of {\it GVs} when the dimension of the Hilbert space is greater than $2$. The following theorem is an adaptation of the KS theorem ---as stated in \cite[Theorem 3.2]{DF}--- to the case of contexts:

\begin{thm}[KS Theorem] If ${\cal H}$ is a Hilbert space of $\dim({\cal H}) > 2$, then a global valuation is not possible.
\end{thm}

KS theorem proves there is no {\it GV}, and thus, an interpretation of projection operators as {\it preexistent} properties becomes problematic. The theory cannot be interpreted, following the classical ontology, as representing a  {\it preexistent} actual state of affairs; i.e., a closed situation considered in terms of one or many systems all of which are definable in terms of a set of preexistent or actual (definite valued) properties. In this case preexistence means existence independent of measurement observations, the properties exist independently of being observed or not. The widespread idea according to which contextuality should be necessarily understood as the claim that `in QM the system behaves differently whether you observe it or not' is simply not related to the KS theorem but to Bohr's ideas and metaphysical presuppositions.\footnote{In fact, this claim is grounded on the strong presuppositions that QM must be necessarily understood in terms of ``common sense'' observation represented in terms of classical notions.} There are four main points which are important to stress with respect to our present analysis: 

\begin{enumerate}

\item[i.]  {\it KS theorem has nothing to do with measurements.} There is no need of actual measurements for the KS theorem to stand. The theorem is not talking about measurement outcomes, but about the possibility (or not) of considering projection operators as preexistent properties. More specifically, the theorem talks about the constraints implied by the formalism to projection operators (interpreted in terms of properties that pertain to a quantum system). Quantum contextuality cannot be tackled through an analysis in terms of measurements simply because there is no reference at all to any measurement outcome. 

\item[ii.] {\it KS theorem is not empirically testable}. As it is well known we cannot measure sets of definite values from a quantum state. In QM we can only measure mean values of observables. The measurement of definite values is restricted to very particular cases.   

\item[iii.] {\it KS theorem makes exclusive reference to the quantum formalism}. KS's notion of contextuality makes reference only to the quantum formalism. In this sense it is an internal formal statement of the theory independent on any particular interpretation of QM.  

\item[iv.]  {\it KS theorem can be read as an ad absurdum proof.} KS makes reference to the interpretation of projection operators in terms of actual (definite valued) properties. Put in a nutshell, KS contextuality deals with the formal conditions that any realist interpretation which respects orthodox Hilbert space QM must consider in order to consistently provide an objective physical representation of reality. 

\end{enumerate}

\noindent To summarize, we can comprise the notion of contextuality implied by KS theorem through the following definition:

\begin{dfn} 
{\bf Kochen-Specker's (Formal-Ontic) Contextuality:} Given a vector in Hilbert, the multiple contexts, considered in terms of bases or complete set of commuting observables, define projection operators which cannot be interpreted as preexistent properties possessing definite values.
\end{dfn}
 
In the following section we will show how the two previous definitions of epistemic and ontic contextuality ---which assume contradictory perspectives--- have been scrambled together into what we have called ``the omelette of quantum contextuality''.

\section{Scrambling KS Theorem with Measurement Outcomes}

While the epistemic viewpoint discusses about measurement processes and outcomes ---assuming that observations are unproblematic---, the ontic perspective is interested in the way the formalism can be conceptually represented beyond observable measurement outcomes, beyond {\it hic et nunc} observations and purely mathematical accounts. In \cite{FuchsPeres00}, Asher Peres made explicit his instrumentalist perspective arguing, together with Christopher Fuchs, that ``quantum theory does not talk about physical reality''; thus, breaking explicitly the link between theoretical representation and {\it physis}. According to their epistemic viewpoint, QM is just ``an algorithm for computing probabilities for the macroscopic events (`detector clicks')''. But, as we shall argue, this perspective becomes highly problematic when addressing the KS theorem. As we have argued above, KS's analysis makes no reference at all to measurement outcomes. On the very contrary, KS theorem discusses the possibility of interpreting the formalism in terms of a particular (meta-)physical representation of reality ---that which conceives existence in (actualist) binary terms. This is the reason why, for anyone attempting to follow an epistemic path with respect to QM, the KS theorem presents an uncomfortable dilemma. 

A radical epistemic proponent might consider a theory to be, following Mach, an ``economy of experience'', following van Fraassen, a formal structure capable of ``saving the phenomena'', or following Peres, ``an algorithm predicting `clicks' in detectors''. Let us concentrate in Peres' instrumentalism. An algorithm provides outputs given an input, but it does not attempt to provide a representation beyond such results. An algorithmic view is one which rejects the need of adding any ontological architecture to the mathematical computation. That is the whole point of being an empiricist, that theories are incapable of representing physical reality. Rather, theories must be regarded as pragmatic ``black boxes'' which help us, humans, to make the right choices ---or in the rephrased view of QBism: it helps us, humans, in choosing what outcome we should bet on \cite{QBism13}.  

But, the KS theorem makes no reference at all to the epistemic realm of outputs (or measurement outcomes). Instead, it reflects about a particular ontological interpretation of the formalism in terms of definite valued preexistent properties. KS theorem is exclusively restricted to a formal and metaphysical analysis. This is an analysis which leaves explicitly aside {\it hic et nunc} observations. KS theorem must be understood as a study about the {\it formal} constraints on projection operators when considered {\it metaphysically} in terms of actual (definite valued) properties. KS explains why the orthodox formalism provides strong restrictions to the possibilities of a classical type interpretation which assumes ---following Newtonian metaphysics--- that the the world is constituted by systems which posses definite valued properties.\footnote{We stress that KS applies only to the orthodox formalism. It does not make reference to different theories like GRW, Bohmian mechanics. Which are not ``interpretations of QM'' but rather different theories which change the mathematical formalism.} In this respect, in order to make our analysis even more clear, we might separate KS theorem in two parts, firstly, KS provides a formal analysis of the mathematical structure of QM in terms of possible binary valuations, secondly, KS uses this purely formal conclusion to provide an {\it ad absurdum} metaphysical proof of the impossibility to understand projection operators in terms of definite valued actual properties. The second part of the theorem makes use of the atomist Newtonian ontology which understands physical reality as represented in terms of systems with definite valued properties. The epistemic viewer might feel quite uncomfortable with the metaphysical discussion presented by the theorem. And she should be worried indeed, for how can the epistemic advocate avoid being dragged into such a purely formal-metaphysical analysis and discussion? A discussion which ---contrary to the epistemic perspective--- makes no reference to measurement outcomes. An analysis in which observation plays no role whatsoever. 

Peres was according to himself a positivist and an instrumentalist.\footnote{In this respect, it is interesting to notice the characterization of Peres of his co-author, Fuchs who explained in a recent article \cite[pp. 4-5]{Fuchs16}:  ``[T]he only person in your present citation list who I would not call a realist is Asher Peres. Asher, in fact, took pride in calling himself alternately a positivist and an instrumentalist. Here are two instances where he labeled himself as such in print: http://arxiv.org/abs/ quant-ph/9711003 and http://arxiv.org/abs/quant-ph/0310010. (You may note that he labelled me with the same terms as well. This was one of the key issues that made the writing of our joint article in {\it Physics Today} [10] so frustrating; every single sentence had to be a careful negotiation in language so that I could feel I wasn't selling my soul.) Asher was fully happy in thinking that the task of physics was solely in making better predictions from sense data to sense data.''} In \cite{Peres91} he provided a formal account of KS theorem followed by an ``intuitive explanation'' of the consequences of the theorem. This ``explanation'' can be found today in almost every paper which discusses about the now famous theorem.

\begin{quotation}
\noindent {\small``The Kochen-Specker (1967) theorem is of fundamental importance for quantum theory. It asserts that, in a Hilbert space of dimension $d$ > 2, it is impossible to associate definite numerical values, 1 or 0, with every projection operator $P_m$, in such a way that, if a set of {\it commuting} $P_m$ satisfies $\sum P_m = 1$, the corresponding values, namely $v(P_m) = 0$ or $1$, also satisfy $\sum v(P_m) = 1$. The thrust of this theorem is that any cryptodeterministic theory that would attribute a definite result to each quantum measurement, and still reproduce the statistical properties of quantum theory, must necessarily be contextual. {\it Namely, if three operators, $A$, $B$ and $C$ satisfy  $[A, B] = [A, C] = 0$ and $[B, C] \neq 0$, the result of a measurement of $A$ cannot be independent of whether $A$ is measured alone, or together with $B$, or together with $C$.}'' [{\it Op. Cit.}, p. 175] (emphasis added)}
\end{quotation}

\noindent As we remarked, this explanation has become extremely popular and is found in many papers which discuss the physical meaning of quantum contextuality as exposed by the KS theorem (e.g. \cite{Appleby05, BarretKent04, Meyer99, Spekkens05}). We can resume this reading in the following definition: 

\begin{dfn} {\sc Epistemic Reading of KS Theorem:} The measurement outcome of the observable $A$, when measured together with $B$ or together with $C$, will necessarily differ in case $[A, B] = [A, C] = 0$, and $[B, C] \neq 0$. 
\end{dfn}

\noindent I myself accepted implicitly this ``reading'' of the KS theorem in a recent paper \cite{deRonde16a}.\footnote{The reason for this is that I myself studied with the book of Peres, {\it Quantum Theory: Concepts and Methods}, when following a course on quantum foundations by Juan Pablo Paz many years ago during my physics degree at the University of Buenos Aires. The book \cite{Peres02} repeats this paragraph almost in its exact original form \cite{Peres91}.} However, when analyzed critically and in detail it is possible to see that, because of the improper mixture of epistemic and ontic presuppositions, the statement has no rigorous physical nor philosophical content. {\sc Definition 5.1} makes reference to an experimental situation which, by definition, cannot be empirically tested. It talks about measurements that cannot be measured, about observations that ---by definition--- cannot be observed! This is what is called an {\it oximoron}. 

KS theorem is not empirically testable in a direct manner, simply because it never talks about measurement outcomes. This is clearly a problem for the epistemic follower since as we know, ``unperformed experiments have no results'' \cite{Peres78}. Were we to assume a consistent epistemic perspective, we should try not to make reference to a metaphysical reality beyond observability. That would be accepting right from the start an ontic perspective regarding the meaning of a theory in terms of the representation of physical reality independent of observations. Something that Peres, as an instrumentalist, wished to avoid at all costs. Such ontic perspective would force us to abandon our epistemic viewpoint according to which a physical theory must be regarded only as an economy of experience, as an algorithm that predicts `clicks' in detectors. KS talks explicitly about the possibilities of interpreting QM. So after all, it does seem that interpreting the mathematical formalism of QM beyond `clicks' in detectors ---contrary to the claim by Fuchs and Peres \cite{FuchsPeres00}--- leads us to very interesting and fruitful conclusions. 

In order to clearly understand our analysis, it is of deep importance to distinguish between the {\it ontic} incompatibility of properties and the {\it epistemic} incompatibility of measurements. The fact that even in classical physics we can find epistemically incompatible measurement situations has been very clearly discussed by Diederik Aerts in \cite{Aerts82}. Aerts discusses the example of a piece of wood which might posses the properties of being ``burnable'' and ``floatable''. Both (classical) properties are {\it testable} through mutually incompatible experimental arrangements. Indeed, in order to test whether the piece of wood can be burned (the ``burnability'' as it were), we must light it up and see if it burns, but then ---because in fact it will burn--- it will no longer be possible to test whether the piece of wood floats. In order to measure the ``floatability'' we must place the piece of wood in a suitable container filled with water and see what happens. The case is, that a burned wood will not float, and also, a wet piece of wood will not burn. Hence, both properties cannot be tested simultaneously. These two experiments are {\it epistemically incompatible}. However, the properties are not {\it ontically incompatible}, the epistemic realm of measurements does not make any direct reference to the ontic level of properties. In classical physics, all properties can be thought to exist as actual (ontic) properties due to the fact that the formal Boolean structure of propositions allows an interpretation in terms of an actual state of affairs. The following two definitions are of  importance to make clear our analysis:

\begin{dfn}
{\sc Epistemic Incompatibility of Measurements:} Two contexts are epistemically incompatible if their measurements cannot be performed simultaneously. 
\end{dfn}

\begin{dfn}
{\sc Ontic Incompatibility of Properties:} Two contexts are ontically incompatible if their formal elements cannot be considered as simultaneously preexistent. 
\end{dfn}

\noindent As we have argued above, while the Bohrian account of contextuality discusses the epistemic incompatibility of measurement situations, KS contextuality addresses the formal incompatibility of projection operators in terms of $GV$ and the impossibility of applying classical ontology to understand the formal structure of the theory.  

Even though classical mechanics might in principle present an epistemic incompatibility of measurements, due to its commutative (Boolean) structure there is never an ontic incompatibility between classical properties. Due to the structure of the (classical) formalism, classical properties are always ontic compatible between each other. On the contrary, in QM the KS theorem makes explicit the ontic incompatibility between properties. This important result is a consequence of the formalism itself. The epistemic incompatibility in QM appears only when classical contexts are considered. But in such case, since KS does not provide a way to test empirically the statement, the discussion becomes completely metaphysical. The epistemic viewer, when entering the KS territory, seems to have been trapped in a metaphysical net with no reference to observable measurement outcomes. 

To conclude, the claim that ``in QM the properties of a system are different whether you look at them or not'' is precluded by KS theorem, since in order to make this claim one requires, as a presupposition, the idea that QM talks in fact about systems in a classical manner ---only then one runs into trouble. On the contrary, KS shows that the very hypothesis that QM talks about classical systems is simply incompatible with the orthodox formalism. 

There is no escape, KS cannot be empirically tested; it makes only reference to the definite valuedness of projection operators, not to measurement outcomes. KS does not provide an empirical result, it is a discussion about the limits imposed by the formalism of QM to the metaphysical mode of existence (binary global valuation) of properties (projection operators). KS makes explicit the deep metaphysical problem that any interpretation of QM must face in case it attempts to interpret the theory in terms of an objective state of affairs. This is why, an ``epistemic reading'' of KS theorem is simply untenable. In conclusion, the KS debate is a purely ontic debate, it has no epistemic elements at play.

\section{Preexistent Properties or Statistical Measurements?} 

The widespread epistemic interpretation of KS theorem which explains the theorem in terms of measurement outcomes has made possible to produce another improper scrambling, namely, that of KS theorem with statistical inequalities of the Boole-Bell type \cite{Bell66, Pitowsky94}. In turn, this has also produced what is known today as ``KS inequalities'' or ``contextual inequalities'' \cite{Cabello98, Cabello08, KunjwalSpekkens15, Spekkens05}.\footnote{Today, it is becoming even common to talk, instead of ``the KS theorem'', of the ``Bell-Kochen-Specker Theorem'' \cite{Appleby05, Cabello98, Kernaghan94, MalleyFine14}. We will leave a detailed analysis of this other improper scrambling for a future work.} To end this paper we would like to discuss an argument which exposes the difficulties ---within the theory of quanta--- of shifting the KS debate from the consideration of the consistency of define values of (preexistent) properties to the discussion about statistical outcomes in a repeated series of measurements. 

Given a vector in Hilbert space $\Psi$ which represents a quantum system it is possible to ask whether the projection operators that arise from different bases can be considered consistently to possess definite values. As we have discussed above, the KS theorem provides a negative answer for dimension higher than 2. Let us remark that KS theorem discusses about a single quantum system at one instant of time. The testability of KS theorem would require the simultaneous measurement of mutually incompatible observables; something which, by definition, is simply impossible within QM. As we discussed in section 5, in QM this is not an epistemic impossibility but rather ---due to the non-commutativity of the algebra of observables--- an ontic impossibility. In QM, to measure all properties of the {\it the same} quantum system one requires {\it necessarily} mutually incompatible measurement setups. This imposes the necessity of a repeated series of measurements. But, is it possible to measure {\it the same quantum system} repeatedly? The answer is well known for quantum physicists: a simple NO. The reason is twofold:

\begin{enumerate}
\item[i.] {\it Measurements in QM are always destructive measurements.} The quantum system is destroyed when measured. If an elementary particle hits a photographic plate we might learn about its energy, its spin, etc., but as a consequence of the imprint the electron will have ceased to exist. 
\item[ii.] {\it Quantum systems cannot be cloned.} The well known ``no-cloning'' theorem \cite{WootersZurek82, Dieks82} of QM shows it is not possible to create a copy of a quantum system.  
\end{enumerate}

\noindent Since measurements in QM are always destructive and quantum systems cannot be cloned, it is not possible to measure a quantum system repeatedly. However, the repeatability of measurements is a necessary condition for statistical analysis of systems. Thus, it is not at all clear what is meant by a ``KS statistical inequality''. What is being truly measured within such ``contextual inequalities''?

The restriction to the measurability of an individual quantum system seems to have not been addressed within the debate about contextual inequalities. We might also remark that the analogy with Boole-Bell type inequalities is not self evident nor direct. In fact, the Bell inequality does not make any direct reference to QM. On the very contrary, it is a statistical statement about classical theories which presuppose the existence of classical probability and classical systems. In more general terms, Bell inequalities presuppose the existence of an actual state of affairs.\footnote{As noticed by Schr\"odinger \cite[p 115]{Bub97} in a letter to Einstein: ``It seems to me that the concept of probability is terribly mishandled these days. Probability surely has as its substance a statement as to whether something {\it is} or {\it is not} the case ---an uncertain statement, to be sure. But nevertheless it has meaning only if one is indeed convinced that the something in question quite definitely {\it is} or {\it is not} the case. A probabilistic assertion presupposes the full reality of its subject.''} Bell inequality is a constrain to the classical modeling of correlations, if the inequality is violated this means that there will exist no classical theory which is capable of reproducing the outcomes. Furthermore, contrary to the quantum case, in the classical case it is always possible to produce repeatable measurements on the same system.\footnote{We leave a more detailed analysis of the distant relation between KS theorem and Bell inequalities for a future work.} This also implies the possibility to bypass the epistemic incompatibility of measurements ---such as those discussed in the previous section. 

The conditions to test KS theorem are not meet for quantum systems. When jumping from the KS discussion about preexistent properties to the debate about statistical measurable inequalities one must fulfill the conditions which secure the tenability of the shift. As it stands, the present debate about KS inequalities seems to have a very insecure foundation. In this respect we might recall what Mermin himself has remarked, namely, that ``the whole notion of an experimental test of KS [theorem] misses the point''.

\section*{Conclusion}
	
In this paper we have argued that there are two different ideas of quantum contextuality within the literature. While the first epistemic notion is due to Bohr's analysis of the measurement process in classical physics and metaphysics, the second formal-ontic definition relates to the intrinsic contextual aspect of the quantum formalism as exposed through the KS theorem. We have shown how these two different understandings of contextuality have been scrambled in the literature through an epistemic misreading of the KS theorem in terms of measurement outcomes. In this respect, we have provided arguments which explain why this widespread understanding of KS contextuality is untenable. Finally, we have also provided arguments which criticize the present discussion in terms of KS ``contextual inequalities'' or the possibility of testing the theorem empirically.  

\begin{center}
\section*{Addendum (to earlier versions of this article)}
\end{center}

\noindent The first version of the present article was submitted to {\it Los Alamos Archives} on the 13th of June 2016. One year later, on a paper submitted on the 27th of July 2017, Karl Svozil showed explicitly the lacunas in the reasoning behind the so called ``contextual inequalities'' which attempt to test experimentally KS theorem. In section 5 of his article \cite{Svozil17}, {\it How can you measure a contradiction?}, he explains the following:  \\

\begin{quotation}
\noindent {\small``Clifton replied with this (rhetorical) question after I had
asked if he could imagine any possibility to somehow ``operationalize''
the Kochen-Specker theorem.

Indeed, the Kochen-Specker theorem ---in particular, not
only non-separability but the total absence of any two-valued
state--- has been resilient to attempts to somehow ``measure''
it: first, as alluded by Clifton, its proof is by contraction ---any
assumption or attempt to consistently (in accordance with admissibility)
construct two-valued state on certain finite subsets
of quantum logics provably fails.

Second, the very absence of any two-valued state on such
logics reveals the futility of any attempt to somehow define
classical probabilities; let alone the derivation of any
Boole's conditions of physical experience ---both rely on,
or are, the hull spanned by the vertices derivable from
two-valued states (if the latter existed) and the respective
correlations. So, in essence, on logics corresponding to
Kochen-Specker configurations, such as the G2-configuration
of Kochen-Specker [90, p. 69], or the Cabello, Estebaranz
and Garc\'ia-Alcaine logic [120, 156] depicted in Fig. 16 which
(subject to admissibility) have no two-valued states, classical
probability theory breaks down entirely ---that is, in the most
fundamental way; by not allowing any two-valued state.

It is amazing how many papers exist which claim to ``experimentally verify'' the Kochen-Specker theorem. However, without exception, those experiments either prove some kind of Bell-Boole of inequality on single-particles (to be fair this is referred to as ``proving contextuality;'' such as, for instance, Refs. [168-172]); or show that the quantum predictions yield complete contradictions if one ``forces'' or assumes the counterfactual co-existence of observables in different contexts (and measured in separate, distinct experiments carried out in different subensembles; e.g., Refs. [156, 173-176]; again these lists of references are incomplete.)'' \cite[sect. 5]{Svozil17}}
\end{quotation}
 
While our article analyzed, from a philosophical perspective, the misleading presuppositions behind the epistemic reading of KS theorem and its consequences for the derivation of the so called ``contextual inequalities'' ---i.e., inequalities supposedly derived from the KS theorem or inequalities which supposedly test the theorem empirically--- Svozil arrived to exactly the same conclusion, now from a logical analysis [{\it Op. cit.}]. I believe this logical confirmation of our previous philosophical investigation shows the importance of critical thought within the foundational debates of QM.
 
As we argued in our paper, the hidden agenda behind the need to develop ``contextual inequalities'' is grounded on the attempt to support an epistemic account of physics. This viewpoint attempts to overcome the fact that KS theorem presents, in itself, a very difficult dilemma for those claiming that the formal mathematical structure of theories can be understood in purely mathematical terms, without any reference to metaphysical presuppositions. Epistemic perspectives of this type are forced in this particular case to either abandon the discussion about KS theorem, or accept that the discussion and analysis of the representation of physical reality must be provided not only in terms of mathematics but also through the formulation of adequate physical concepts. KS theorem in its complete formal-ontological form can be then understood as an {\it ad absurdum} proof of the untenability of the classical metaphysical presupposition according to which reality must be necessarily represented in terms of classical ontology, namely, as systems with definite valued preexistent properties, or in more general terms, as an actual state of affairs.

\section*{Acknowledgments} 

I want to thank Karl Svozil for explaining to me the logical error in the derivation of the so called ``contextual inequalities''. I also want to thank Kevin Vanslette for discussions on earlier versions of this article. This work was partially supported by the following grants: FWO project G.0405.08. CONICET RES. 4541-12 and Project PIO-CONICET-UNAJ (15520150100008CO) ``Quantum Superpositions in Quantum Information Processing''.


\end{document}